\documentclass[usenatbib]{emulateapj}
\usepackage{natbib}
\newcommand{\lta}{\la}

\newcommand{\kpc}{\>{\rm kpc}}

\begin{document}
\title{Distribution of satellite galaxies in high redshift groups}

\author{Yougang Wang\altaffilmark{1}, Changbom Park\altaffilmark{2},
Ho Seong Hwang\altaffilmark{2,3},Xuelei Chen\altaffilmark{1,4}}

\altaffiltext{1}{Key Laboratory of Optical Astronomy,
National Astronomical Observatories, Chinese
Academy of Sciences, Beijing 100012, China; E-mail:
wangyg@bao.ac.cn}

\altaffiltext{2}{School of Physics, Korea Institute for Advanced
Study, Dongdaemun-gu, Seoul 130-722, Korea}

\altaffiltext{3}{CEA Saclay/Service d'Astrophysique, F-91191
Gif-sur-Yvette, France}

\altaffiltext{4}{Center of High Energy Physics, Peking University,
Beijing 100871, China}

\begin{abstract}
We use galaxy groups at redshifts between 0.4 and 1.0 selected from
the Great Observatories Origins Deep Survey (GOODS) to study the
color-morphological properties of satellite galaxies, and
investigate possible alignment between the distribution of the
satellites and the orientation of their central galaxy. We confirm
the bimodal color and morphological type distribution for satellite
galaxies at this redshift range: the red and blue classes
corresponds to the early and late morphological types respectively,
and the early-type satellites are on average brighter than the
late-type ones. Furthermore, there is a {\it morphological
conformity} between the central and satellite galaxies: the fraction
of early-type satellites in groups with an early-type central is
higher than those with a late-type central galaxy. This effect is
stronger at smaller separations from the central galaxy. We find a
marginally significant signal of alignment between the major axis of
the early-type central galaxy and its satellite system, while for
the late-type centrals no significant alignment signal is found. We
discuss the alignment signal in the context of shape evolution of
groups.

\end{abstract}


\maketitle

\section{Introduction}
In a cold dark-matter dominated universe, galaxies form within dark
matter halos, and smaller halos form first, subsequently these may
grow larger by accreting material and/or by merging with other
halos. As a result, satellite galaxies are distributed within the
dark matter halo of galaxy groups. Under the assumption that there
is an unbiased distribution between the satellites and dark matter
halo, the position of satellites can be used to determine the shape
of the dark matter halo \citep{1980MNRAS.191..325C,
1991MNRAS.249..662P,1993ApJ...416..546F,2000MNRAS.316..779B,
2001MNRAS.325..133O,2004MNRAS.352.1323P,2006ApJ...650..770P,
2008MNRAS.385.1511W}, and their kinematics could be used to estimate
the halo mass
\citep{1993ApJ...405..464Z,1997ApJ...478...39Z,2002ApJ...571L..85M,
2003ApJ...593L...7B,2004ApJ...600..657K,2004MNRAS.352.1302V,
2009MNRAS.392..917M}. Or, if a spatial bias between the distribution
of satellites and the underlying dark matter distribution is found,
it would be a very important clue for us in the study of galaxy
formation theory. The orientation of the satellites may also provide
useful information on its formation and evolution process.
High-resolution simulations have shown that subhalos tend to align
with the major axis of their host halos
\citep{2004ApJ...603....7K,2008MNRAS.386L..52K,2008MNRAS.388L..34K,
2005MNRAS.363..146L,2005MNRAS.364..424W,2005ApJ...629..219Z,2006ApJ...644L..25A,
2007MNRAS.378.1531K,2008ApJ...675..146F,2010arXiv1002.2853K}. Such
effects could be examined with observations of satellites at
different redshifts, to give a more comprehensive test of the
theoretical model.

The morphology and color of the satellites are directly related to
formation history of the host group. It has long been known that
galaxies exhibit a bimodality in color and morphology:
morphologically early-type galaxies which are typically red and have
little or no ongoing star formation, and morphologically late-type
galaxies, typically blue with active star formation. It is well
known that galaxy morphology depends on local density environment.
\cite{1931ApJ....74...43H} found a larger population of ellipticals
and lenticulars in galaxy clusters, and subsequent studies revealed
the connection between galaxy morphology and environment in low
redshift clusters of galaxies
\citep{1974ApJ...194....1O,2006MNRAS.366....2W,2009ApJ...699.1595P},
nearby galaxy pairs \citep{2007ApJ...658..898P,2008ApJ...674..784P},
isolated galaxy-scale satellite systems \citep{2008MNRAS.389...86A},
and galaxy pairs at high redshifts \citep{2009ApJ...700..791H}. The
color bimodality is noted by numerous studies at both low redshifts
\citep{2001AJ....122.1861S,2003ApJ...594..186B,
2004ApJ...600..681B,2004MNRAS.353..713K} and at high redshifts of
$z\sim 1$
\citep{2004ApJ...608..752B,2005MNRAS.362..268T,2005ApJ...620..595W,
2007MNRAS.381..962C}. However, investigation on the bimodality and
morphology-radius relation for satellite galaxies in high redshift
groups have so far been lacking, one of the aims of the present
paper is to investigate these issues.

The alignment between satellite distribution and the orientation of
the central galaxy has been studied extensively since the work of
\cite{1969Ark.Astron....5..305}, who first found that satellites are
located preferentially close to the minor axes of the central
galaxy. Many subsequent works confirmed the presence of this
``Holmberg effect''
\citep{1976MNRAS.174..695L,1982Observatory.102..202,1994ApJ...431L..17M,
1996ASPC...92..444H,1997ApJ...478...39Z,2000AJ....119.2248H,2005A&A...431..517K,2006AJ....131.1405K,
2006MNRAS.365..902M,2007MNRAS.374.1125M}, though not all reached the
same conclusion
\citep{1975AJ.....80..477H,1979MNRAS.187..287S,1982MNRAS.198..605M}.
One common limitation of the early studies is that the sample used
are relatively small. With the advent of the 2dF Galaxy Redshift
Survey (2dFGRS) \citep{2001MNRAS.328.1039C} and the Sloan Digital
Sky Survey (SDSS) \citep{2000AJ....120.1579Y}, much larger samples
became available. \cite{2004MNRAS.348.1236S,2009MNRAS.395.1184S}
found a tendency for the satellites to be located along the host
major axes at the large-scale of $300 \kpc \lta r_p \lta 500 \kpc$
for a set of 1489 host galaxies with 3079 satellites from the
2dFGRS. This result is in a good agreement with similar studies
carried out on the SDSS data
\citep{2005ApJ...628L.101B,2007MNRAS.376L..43A,
2008MNRAS.390.1133B,2010ApJ...709.1321A}. \cite{2006MNRAS.369.1293Y}
and \cite{2008MNRAS.385.1511W} found a strong dependence of the
alignment signal on the color of the central and satellite galaxies
using groups in the SDSS Data Release 2 (DR2) and Data Release 4
(DR4) respectively. These studies found that the alignment signal is
strongest between red central galaxies (hereafter `centrals') and
red satellites, while the satellites of blue centrals were
consistent with being distributed isotropically. The alignment
strength is also a function of the group mass, with stronger
alignment signal in more massive groups. Other forms of alignment
have also been studied, and some significant signals are detected.
These include the alignment between neighboring clusters
\citep{1982A&A...107..338B,1989ApJ...344..535W,1994ApJS...95..401P,2009ApJ...703..951W},
between brightest cluster galaxies (BCGs) and their parent clusters
\citep{1980MNRAS.191..325C,1982A&A...107..338B,1990AJ.....99..743S,
2010arXiv1003.0322N}, between the orientation of satellite  galaxies
and the orientation of the cluster
\citep{1985ApJ...298..461D,2003ApJ...594..144P}, and between the
orientation of satellite galaxies  and the orientation of  the BCG
\citep{1990AJ.....99..743S,2007ApJ...662L..71F}.

For obvious reasons, most studies on satellite distribution have
been limited to
low redshifts, typically $z\lta0.2$. Recently,
\cite{2006MNRAS.369..479D} analysed a sample of relatively high redshift
($0.4<z<0.5$) Luminous Red Galaxies (LRGs) extracted from the
SDSS DR4 and their surrounding structures to explore the presence of
alignment effects. They confirmed that such alignment effect
was also present at $z\sim0.5$.
\cite{2009ApJ...694..214O} investigated the correlation between the
orientation of giant ellipticals by measuring the intrinsic
ellipticity correlation function of 83,773 SDSS LRGs at redshifts
0.16-0.47 and also found a positive alignment
between pairs of the LRGs up to  $~30h^{-1}$Mpc scales.

Deep galaxy redshift surveys such as the Great
Observatories Origins Deep Survey (GOODS)
\citep{2004ApJ...600L..93G}
%and the Deep Extragalactic EvolutionaryProbe 2 (DEEP2) \citep{2003SPIE.4834..161D,2005ASPC..339..128D}
now enable us to study the satellite distribution at even higher
redshifts. Another aim of our paper is to detect the alignment
between the distribution of satellite galaxies and the orientation
of their central galaxy at redshifts beyond the local universe by
using GOODS data. We construct a group catalog using a
Friends-of-Friends (FoF) method, then we study the alignment signals
with this sample. Compared with the previous studies, the redshift
range in our sample is $0.4\leqslant z\leqslant1.0$, which includes
many high redshift groups ($z\sim 1$).

This paper is organized as follows. We describe the
observational data used for this study in section 2, and the FoF
group-finding method in section 3. In section 4, we present the properties of
satellites in groups. The method to quantify the alignment signal
is presented in section 5, and the results given in section 6. Section 7
summarize our results and discuss various related issues.

\section{Observational data set}
\subsection{galaxy sample}
The galaxy sample used here is selected by
\cite{2009ApJ...700..791H}. Here we give a brief description of the
data sample, and we refer the reader to Hwang \& Park for more
details.

We used a spectroscopic sample of galaxies in GOODS. GOODS is a deep
multiwavelength survey covering two carefully selected regions
including the Hubble Deep Field North (HDF-N, hereafter GOODS-North)
and the Chandra Deep Field South (CDF-S,hereafter GOODS-South).
Total observing area is approximately 300 $\rm{arcmin}^2$ and each
region was observed by NASA's Great Observatories ({\it HST}, {\it
Spitzer} and {\it Chandra}), ESA's {\it XMM-Newton}, and several
ground-based facilities. {\it HST} observations with Advanced Camera
for Surveys (ACS) were conducted in four bands: $B$ (F435W, 7200s),
$V$ (F606W, 5000s), $i$ (F775W, 5000s), and $z$ (F850LP, 10,660s).
Among the sources in the ACS photometric catalog,
\cite{2009ApJ...700..791H} selected 4443 (2197 in GOODS-South, 2246
in GOODS-North) galaxies whose reliable redshifts are available. In
our analysis, a volume-limited sample of 1332 galaxies with
$M_B\leq-18.0$ and $0.4\leq z\leq1.0$ is used. The rest-frame
$B$-band absolute magnitude $M_B$ of galaxies is computed based on
the ACS photometry with Galactic reddening correction
\citep{1998ApJ...500..525S} and K-corrections
\citep{2007AJ....133..734B}. The evolution correction (an increase
of $1.3M_B$ per unit redshift) was also applied to the rest-frame
$M_B$ \citep{2007ApJ...665..265F}.

\subsection{Morphology Classification}
\cite{2009ApJ...700..791H} visually inspected the individual $Bviz$
band images and $Bvi$ color images of the galaxies in a
volume-limited sample. The galaxy is divided into two morphological
types: early types (E/S0) and late types (S/Irr). Early-type
galaxies are those with little fluctuation in the surface brightness
and color and possess good symmetry in morphology, while late-type galaxies show
internal structures and/or variations in the color
images.

\section{Groups of Galaxies}
A very important step in our investigation is to identify the galaxy
groups. There are many different techniques to identify groups in
the local and distant Universe
\citep{2005MNRAS.356.1293Y,2007ApJ...660..239K,
2008AJ....135..809L,2009ApJS..183..197W}. All of these methods have
their own advantages and disadvantages, which we do not discuss
here. We use the FoF method to find the groups. The FoF algorithm
adopted here is that of \cite{2004MNRAS.348..866E} and
\cite{2009ApJ...697.1842K}. There are three adjustable parameters in the
FoF algorithm: the linking length $b$, the maximum perpendicular
linking length in physical coordinates $L_{\rm max}$ and the ratio
between the linking length along and perpendicular to the line of
sight $R$. If two galaxies $i$ and $j$ with comoving distances
$d_{ci}$ and $d_{cj}$ satisfy the following two conditions, then
they are assigned to the same group. The two conditions,
respectively, are
\begin{eqnarray}
\theta_{ij} \leq \frac{1}{2}
\left(\frac{l_{\perp,i}}{d_{ci}}+\frac{l_{\perp,j}}{d_{cj}}\right)
\end{eqnarray}
and
\begin{eqnarray}
\left|d_{ci}-d_{cj}\right| \leq
\frac{l_{\parallel,i}+l_{\parallel,j}}{2}.
\end{eqnarray}
where $\theta_{ij}$ is the angular separation between galaxy $i$ and
$j$, and the two parameters $l_{\perp}$ and $l_{\parallel}$ are the
comoving linking lengths perpendicular and parallel to the line of
sight defined by
\begin{eqnarray}
l_{\perp} &=& \min \left[L_{\rm max}(1+z),\frac{b}{\bar{n}^{1/3}}\right]\\
l_{\parallel} &=& R\; l_{\perp},
\end{eqnarray}
where $L_{\rm max}$ is the maximum perpendicular linking length in
physical coordinates and $\bar{n}$ is the mean density of galaxies.
Since we have a volume-limited sample of galaxies, it is easy to
obtain the value $\bar{n}$. For the typical value of three free
parameters $b$, $L_{\rm max}$ and $R$, we adopted them as listed in
Table 1 of \cite{2009ApJ...697.1842K}.

\begin{figure}
\resizebox{\hsize}{!}{\includegraphics{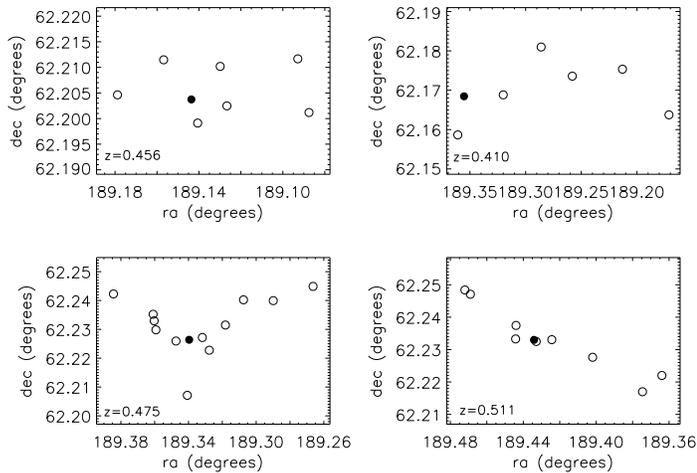}} \caption{Four
typical groups found by the FoF method. The three parameters $b$,
$L_{\rm max}$ and $R$ are set $b=0.11$, $L_{\rm max}=0.45 {\rm Mpc}
$, and $R=13$, respectively. The filled and open circles represent
the central and satellite galaxies, respectively. } \label{group}
\end{figure}

Figure~\ref{group} shows a few groups found by the FoF method. The
three parameters $b$, $L_{\rm max}$ and $R$ are set as $b=0.11$,
$L_{\rm max}=0.45 {\rm Mpc} $, and $R=13$, respectively. In each
group, the filled and open circles represent the central and
satellite galaxies, respectively. The brightest galaxy in each group
is defined as the central galaxy, and the rest are called its
satellites. As shown in the top right panel of Figure~\ref{group},
in some cases the ``central galaxy'' defined by this way is not
actually located in the central region of groups. We call them the
Deviation-Center-Group (DCG), and the other groups
Located-Center-Group (LCG). The criterion for LCG is
$d_a\leq\frac{1}{4}g_a$, where $d_a$ is the angular distance between
the brightest galaxy and the geometrical center of the group, and
$g_a$ is the angular size of the group. The half angular size of the
group is defined as the angular separation between the geometrical
center of the group and farthest member galaxy. In the determination
of the alignment signal, only the LCGs are used. A total of 206
groups are found in our volume-limited sample.

\begin{figure}
\resizebox{\hsize}{!}{\includegraphics{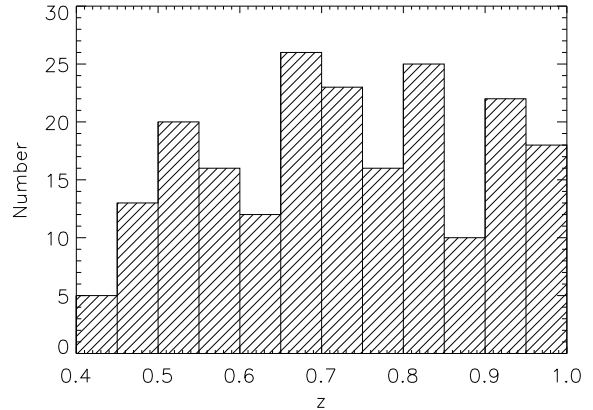}}
\caption{Redshift distribution of the groups.} \label{z_group}
\end{figure}

In Figure~\ref{z_group}, we show the redshift distribution of the
groups in our sample. The distribution has a broad peak near
$z\sim0.7$.

\section{Properties of satellites in groups}
\begin{figure}
\resizebox{\hsize}{!}{\includegraphics{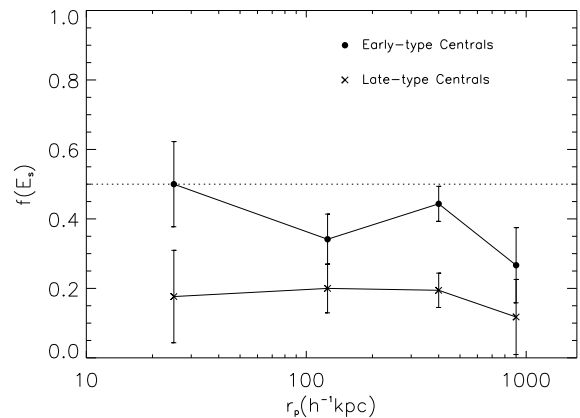}}
\caption{Early-type fraction of satellites galaxies as a function of
projected distance from the central galaxy. The filled circles and
crosses correspond to the early- and late-type central galaxy cases,
respectively.} \label{fra_el}
\end{figure}

In Figure~\ref{fra_el}, we present the early-type fraction of
satellites as a function of projected distance $(r_{\rm p})$ from
the central galaxy in groups. The filled circles and crosses are for
the early- and late-type galaxy cases, respectively. The error bars
represent $68\%$ $(1\sigma)$ confidence intervals, which are
determined with the bootstrap resampling method. It is seen that the
early-type fraction of satellites increases as satellites approach
their early-type central galaxy. Also, the early-type fraction of
satellites in groups with early-type centrals is higher than those
with late-type centrals, which indicates that the satellite galaxies
tend to have morphology similar to their centrals. Similar results
have been obtained for galaxies in groups and clusters by
\cite{2006MNRAS.366....2W}, for galaxy pairs in general environment
by \cite{2007ApJ...658..898P,2008ApJ...674..784P} in low redshift
samples, and by \cite{2007ApJS..172..284C} and
\cite{2009ApJ...700..791H} in high redshift samples. In the group
with a late-type central galaxy, the early-type fraction of
satellites is nearly constant as the distance from the central
galaxy changes.

\begin{figure}
\resizebox{\hsize}{!}{\includegraphics{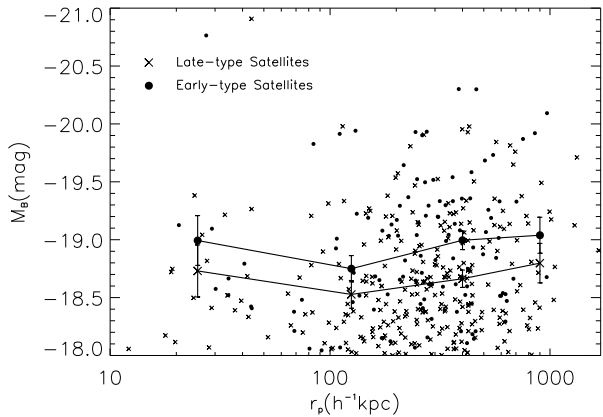}} \caption{Rest
frame $B$-band absolute magnitude $M_B$ of the early- (filled
circle) and late-type (cross) satellites as a function of the
projected distance from the central galaxy. Two solid lines
represent the median values.} \label{Mb_rp}
\end{figure}

In Figure~\ref{Mb_rp}, we show the rest frame $B$-band absolute
magnitude $M_B$ of the early- (filled circle) and late-type (cross)
satellites as a function of the projected distance from the central
galaxy. Two solid lines represent the median values.
The early-type satellites are on the average brighter than the  late-type
satellites.

\begin{figure}
\resizebox{\hsize}{!}{\includegraphics{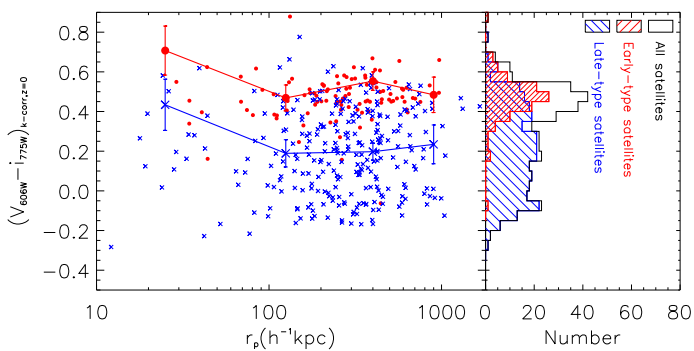}} \caption{Color
of the early- (filled circle) and late-type (cross) satellites as a
function of the projected distance from the central galaxy (left
panel). Two solid lines represent the median values. Color
distributions for the satellite are shown by histograms in the right
panel.} \label{color_rp}
\end{figure}

In the left panel of Figure~\ref{color_rp}, we present the color of
the early- (filled circle) and late-type (cross) satellites as a
function of the projected distance from the central galaxy.
$V_{606W}$ and $i_{775W}$ of galaxies are K-corrected (to z=0)
magnitudes \citep{2007AJ....133..734B}. Two solid lines represent
the median values. In the right panel of Figure~\ref{color_rp}, the
distribution of color of satellites is shown. It can be clearly seen
that the early- and late-type satellites occupy the red and blue
bumps of the bimodal color distribution, respectively.

\section{Quantifying the alignment}
In order to quantify the alignments of objects, we follow the method
in \cite{2005ApJ...628L.101B} and compute the distribution functions
of the alignment angles, $P(\theta)$, where $\theta$ is the angle
between the major axis of the central group galaxy and the direction
of a satellite relative to the centrals. The angle $\theta$ is
constrained in the range $0^{\circ}\leq\theta\leq90^{\circ}$, where
$\theta=0^{\circ}(90^{\circ})$ suggests that the satellite lies
along the major (minor) axis of the central galaxy.

For a  given  set  of  the central and satellite galaxies,  we first
count the total number of central-satellite pairs, $N(\theta)$, for
a number of bins in $\theta$. Next, we construct 200 random samples
in which we randomize the orientations of all centrals, and compute
$\langle N_R(\theta) \rangle$, the average number of
central-satellite pairs as function of $\theta$. The random samples
constructed this way suffer exactly  the same selection effects as
the  real sample, so any significant difference between $N(\theta)$
and $N_R(\theta)$ reflects a genuine alignment between the
orientations of the centrals and the distributions of their
corresponding satellite galaxies.

Following \cite{2006MNRAS.369.1293Y} and
\cite{2008MNRAS.385.1511W},  we  introduce
the distribution of normalized pair counts:
\begin{equation}\label{eq:fpairs}
f_{\rm pairs}(\theta)=\frac{N(\theta)}{\langle N_R(\theta)\rangle}.
\end{equation}
In the  absence of any  alignment, $f_{\rm pairs}(\theta)=1$, while
$f_{\rm pairs}(\theta) >  1$ at small $\theta$ implies a satellite
is preferentially aligned along the major axis of their central
galaxy.

We quantify  the fluctuation  using $\sigma_{\rm  R}(\theta) /
\langle N_{\rm R}(\theta) \rangle$, where $\sigma_{\rm R}$ is the
standard deviation of $N_{\rm R}(\theta)$, and is estimated from the
200 random samples. In addition to  this normalized pair count, we
also compute the average angle $\langle \theta \rangle$. In   the
absence of   any   alignment $\langle\theta\rangle = 45^{\circ}$. If
one finds $\langle \theta \rangle <45^{\circ}$ ($\langle \theta
\rangle
> 45^{\circ}$), it means that the satellites are distributed along
the major (minor) axis of the central galaxy.

\section{Alignment Measurement}
In order to study the alignment signal, the position angles of the
central galaxies are required. We only use those groups with
central galaxy axis ratio $b/a<0.8$. For these galaxies,
the isophotal position angle is well defined. Here $a$ and $b$ are
the isophotal semi-major and minor axis lengths adopted from the
$i$-band measurements in the HST/ACS photometric catalog,
respectively. Finally, we have 168 central-satellite pairs in the
detection of the alignment signal.

\begin{figure}
\resizebox{\hsize}{!}{\includegraphics{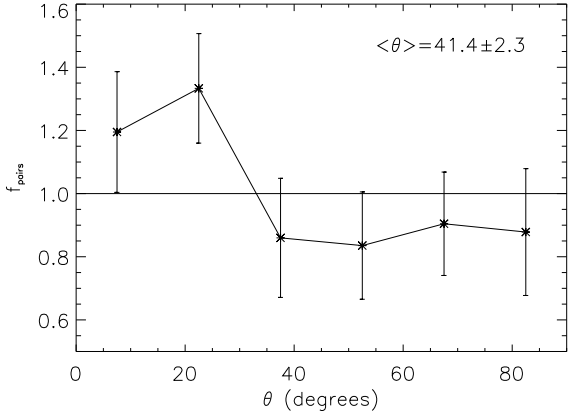}}
\caption{Normalized probability distribution, $f_{\rm
pairs}(\theta)$, of the angle $\theta$ between the major axis of the
central galaxy and the distribution of satellites.} \label{fp_all}
\end{figure}

Figure~\ref{fp_all} shows $f_{\rm pairs}$ for the selected
central-satellite systems. There is a marginally significant signal
of alignment between the orientation of the central galaxies and the
distribution of the satellites. Satellite galaxies are distributed
preferentially along the major axis of their central galaxy. This is
also supported by the fact that $\langle \theta\rangle =
41^{\circ}.4\pm2^{\circ}.3$, which deviates from the case of no
alignment (i.e.$\langle \theta\rangle = 45^{\circ}.0$) by
$1.6\sigma$. Moreover, a Kolmogorov-Smirnov (KS) test also suggests
that an isotropic distribution of satellites in our sample is
rejected with a confidence level higher than $90\%$. If we remove
groups with $z>0.6$ in our sample, then the alignment strength is
changed from $1.6\sigma$ to $1.2\sigma$.

\begin{figure}
\resizebox{\hsize}{!}{\includegraphics{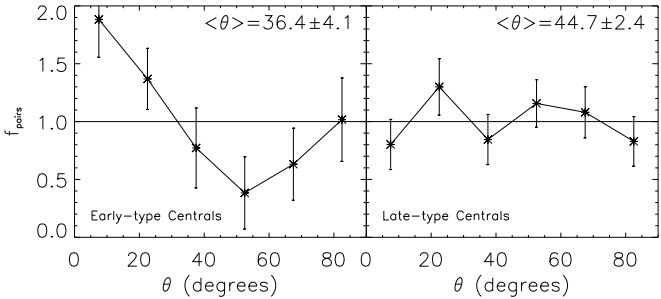}} \caption{Same as
Figure~\ref{fp_all}, but for different subsamples, divided by the
morphological type of the central galaxy of the group.} \label{fp_c}
\end{figure}

In order to study how the alignment depends on the central galaxy
properties, we divide our sample into early-type central and
late-type central cases. Figure~\ref{fp_c} shows the alignment
signals $f_{\rm pairs}(\theta)$ for the sample with early- (left
panel) and late-type (right panel) central galaxies. As can be seen,
systems with an early-type central galaxy shows $2\sigma$ alignment
signal. A KS test finds that the sample with early-type centrals is
not isotropic with confidence level higher than $99\%$.
The distribution of $f_{\rm pairs}(\theta)$ has an
interesting shape, being greater at both $\theta\sim 0^{\circ}$
and $\theta\sim 90^{\circ}$ than at $\theta\sim 45^{\circ}$. This
could perhaps be explained by the effect of
infall from perpendicular filaments on
forming galaxies \citep{2008ApJ...689..678B}. Systems with a
late-type central galaxy, however, show no significant alignment.
%This is
%because a perfect alignment of satellites around a prolate
%early-type central galaxy is when the satellites are distributed
%along the major axis. But a perfect alignment around a disk central
%galaxy is when satellites are distributed on the disk plane. The
%$f_{\rm pairs}(\theta)$ statistic will tell us the system around the
%disk galaxy is less aligned on average depending on the inclination
%of the system.
Compared with the alignment signal detected by
\cite{2006MNRAS.369.1293Y}($\langle \theta\rangle =
42^{\circ}.2\pm0^{\circ}.3$) and \cite{2008MNRAS.385.1511W}($\langle
\theta\rangle = 42^{\circ}.46\pm0^{\circ}.12$) in low redshift
groups, there is no significant difference (the confidence level for
this tiny difference is below $0.5\sigma$) between the alignment
strength in high-z and local groups (for the total samples). In
other words, no evolution of the alignment is seen within redshift
[0,1].

\section{Summary and Discussion}
Using the FoF group finder, we create a high-redshift ($0.4\leq
z\leq1.0$) group catalogue out of a spectroscopic sample of galaxies
in the GOODS fields. We also identify the morphology of the
satellite galaxies visually. The morphologically early- and
late-type satellites occupy the red and blue bumps of the bimodal
color distribution, respectively. We then study the early-type
fraction, the magnitude-radius relation and the color-radius
relations of the satellite galaxies in these groups. We find that
the early-type fraction of satellites in early-host centrals is
higher than those in late-type centrals. The early-type satellites
are also on the average brighter than the late-type satellites.

We measured the alignment between the distribution of satellites and
the orientation of their central galaxy. We find a marginally
significant alignment signal for the whole sample and for the
subsample with early-type centrals. However, we do not find any
alignment signal for the subsample with late-type centrals.

It is known that the group catalog strongly depends on the three
adjustable parameters $b$, $L_{\max}$ and $R$ in the FoF algorithm. In
order to study how these parameters affect the measurement of the
alignment signal, we have adopted five groups of parameter-sets as
listed in Table 1 of \cite{2009ApJ...697.1842K} in identifying groups,
and calculated alignment signals for each case. We find that the result of the
alignment signal from the groups found by using the five different sets
of group parameters are nearly the same. This indicates that our
results on group central-satellites alignment is not sensitive to the
choice of adjustable parameters in the FoF algorithm.

The measured alignment signal may be compared with numerical simulation
results. Using N-body simulations,
\cite{2002ApJ...574..538J,2004ApJ...617..847W}
found that the non-sphericity of dark matter
halos are greater at higher redshift, so we might expect a
stronger alignment strength in the high-redshift groups.
However,  in our high redshift catalog we only find  an
alignment strength similar to the local groups.

There are several possibilities for this result.
First, due to the limited number of
pairs available, the sample variance is still large, and the detection is
marginal. It is still difficult to draw conclusions from this observation.
We have 168 central-satellite pairs in total
and find a 1.6$\sigma$ alignment signal.
To check how the strength of alignment signal depends on the sample size,
we use the SDSS DR4 data adopted in \cite{2008MNRAS.385.1511W} to make a
test. The total number of central-satellite pairs in
\cite{2008MNRAS.385.1511W} is 62212, and the alignment signal is
21$\sigma$($\langle \theta\rangle = 42^{\circ}.46\pm0^{\circ}.12$).
We randomly select 168, $168\times2$, $168\times4$ central-satellite
pairs from this large sample, the corresponding alignment
signal is detected at
$2.4$, $2.5$ and $3.9\sigma$, respectively. If the same scaling is applicable
to the high redshif sample, we need to
increase our sample size by at least fourfold to reach
a significant detection of about $3\sigma$.

Second, we have assumed that stronger non-sphericity produce
stronger alignments. To certain extent this should be true, as there
would be no alignment signal when the distribution is spherical.
However, at different redshifts the relation between non-sphericity
and alignment may be different. For the same non-sphericity, the
alignment might be weaker at higher redshifts due to some reason
(e.g., less time for dynamical adjustment), thus partly compensated
for effects of the stronger non-sphericity.
% {\bf It
%has been noted that the ellipticity is larger and the alignment signal is
%stronger for more massive halos in low redshift groups by
%\cite{2008MNRAS.385.1511W}, but we are not clear whether this
%shape-alignment relation is valid in high readshift samples. ??}

Finally, the predictions by \cite{2002ApJ...574..538J,2004ApJ...617..847W}
were based on N-body simulations.
Inclusion of baryon cooling effect may affect the shape of the
halo \citep{2004ApJ...611L..73K,2008ApJ...681.1076D}, and change
the conclusions on the shape evolution of dark matter halos.

We may also compare the observational results of
alignment strength at different redshifts in the literature.
For the group in the range $0.01\leq z\leq0.2$ \citep{2008MNRAS.385.1511W},
the misalignment angle between the major axis of the central galaxy
and the projected major axis of the host halo follows a
Gaussian distribution with zero mean and a dispersion of $~23^{\circ}$. Using a
sample at $(0.16\leq z\leq0.47)$,
\cite{2009ApJ...694..214O} discovered that the misalignment
angle between the central LRGs and their host halo also
follows a Gaussian distribution with a zero mean, but the dispersion
angle is $\sim35^{\circ}$. If we extend this trend
to samples at even higher redshifts,
the misalignment angles should be larger. With such a trend, a weaker
alignment signal is expected at higher redshifts.

In order to further improve our
understanding of the spatial distribution of satellites in high
redshift groups, we need a large sample. Some undergoing surveys, such
as the Canada-France-Hawaii Telescope (CFHT) Legacy Survey, may
extend the sample size significantly,
and bring a definite answer to questions related to
the distribution of satellites in the high redshift groups.

\section*{Acknowledgments}
We sincerely thank the referee for the constructive and detailed
comments and suggestions. This  work has started during  YGW's visit
to KIAS,  and he would like to express  his gratitude to KIAS. CBP
acknowledges the support of the Korea Science and Engineering
Foundation (KOSEF) through the Astrophysical Research Center for the
Structure and Evolution of the Cosmos (ARCSEC). YGW acknowledge the
support by the Young Researcher Grant of National Astronomical
Observatories. XLC acknowledge the support by the NSFC Distinguished
Young Scholar Grant No.10525314. YGW and XLC are also supported by
the Ministry of Science and Technology under the 973 program
(2007CB815401, 2010CB833004), and the CAS Knowledge Innovation
Program  (Grant No. KJCX3-SYW-N2).

\bibliography{high-z}
\bibliographystyle{apj}

\end{document}